\documentclass[newstyle,twocolumn,proceedings]{rmaa}

\usepackage{rmaacite}

\renewcommand{\P}[1]{%
\ifnum#1=1\hbox{OW~168--326E}\fi
\ifnum#1=2\hbox{OW~167--317}\fi
\ifnum#1=3\hbox{OW~163--317}\fi
\ifnum#1=5\hbox{OW~158--323}\fi
\ifnum#1=0\hbox{OW~171--334}\fi}

\title{New Results from a Survey of Galactic Outflows in Nearby Active
Galactic Nuclei}
\author{S. Veilleux\altaffilmark{1}, G. Cecil\altaffilmark{2}, J. Bland-Hawthorn\altaffilmark{3}, P. L. Shopbell\altaffilmark{4}}
\altaffiltext{1}{Department of Astronomy, University of Maryland, College Park, Maryland}
\altaffiltext{2}{Department of Physics and Astronomy, University of North Carolina, Chapel Hill, North Carolina}
\altaffiltext{3}{Anglo-Australian Observatory, Epping, Australia}
\altaffiltext{4}{Department of Astronomy, California Institute of Technology, Pasadena, California }

\fulladdresses{
\item S. Veilleux: Department of Astronomy, University of Maryland, College Park, MD 20742 (veilleux@astro.umd.edu)
\item G. Cecil: Department of Physics and Astronomy, University of North Carolina, Chapel Hill, NC 27599-3255. (cecil@physics.unc.edu)
\item J. Bland-Hawthorn: Anglo-Australian Observatory, P.O. Box 296 Epping, NSW 2121 Australia. (jbh@aaoepp2.aao.gov.au)
\item P. L. Shopbell: Department of Astronomy, California Institute of Technology, Pasadena, CA 91125. (pls@astro.caltech.edu)
}

\shortauthor{Veilleux et al.}
\shorttitle{Galactic Outflows in Nearby AGNs}


\abstract{%
Recent results from a multiwavelength survey of spatially resolved
outflows in nearby active galaxies are presented. Optical Fabry-Perot
and long-slit spectroscopic data are combined with VLA and ROSAT
images, when available, to probe the warm, relativistic and hot gas
components involved in the outflow. The emphasis is put on
objects which harbor wide-angle galactic-scale outflows but also show
evidence at radio or optical wavelengths for collimated jet-like
phenomena (e.g., Circinus, NGC 4388, and to a lesser extent NGC
2992). Our results are compared with the predictions from published
jet-driven thermal wind models. }

\listofauthors{S.~Veilleux, G.~Cecil, J.~Bland-Hawthorn, P.~L.~Shopbell}
\indexauthor{Veilleux, S.}
\indexauthor{Cecil, G.}
\indexauthor{Bland-Hawthorn, J.}
\indexauthor{Shopbell, P.~L.}

\begin{document}

\maketitle

\section{Introduction}

The main topic at this conference is {\bf jet}-entrained material in
Herbig-Haro (HH) objects and in active galactic nuclei (AGN).
However, as we discuss in this paper, a significant fraction of AGNs
harbor poorly collimated winds which may also be of great dynamical
significance. A similar wind phenomenon is known to take place in a
number of HH objects (e.g., HH 111; Nagar et al. 1997).

A broad range of processes may be responsible for wide-angle outflows
in AGNs:
\begin{itemize}
\item[1.] {\bf Starburst-driven winds.} The deposition of a large
amount of mechanical energy by a nuclear starburst may create a
large-scale galactic wind -- superwind -- which encompasses much of
the host galaxy. Depending upon the extent of the halo and
its density and upon the wind's mechanical luminosity and duration,
the wind may ultimately blow out through the halo and into the
intergalactic medium (e.g., Stickland \& Stevens 2000, and references
therein). The outflow is expected to be wide-angled and oriented roughly
along the minor axis of the galactic disk.
\item[2.] {\bf X-ray--heated torus winds.} X-rays emitted in the inner
part of an accretion disk can Compton-heat the surface of the disk
further out, producing a corona and possibly driving off a wind (e.g.,
Begelman, McKee, \& Shields 1983; Krolik \& Begelman 1986, 1988;
Balsara \& Krolik 1993). For Seyfert galaxies, a substantial wind with
T$_{\rm wind}$ $\approx$ 1 $\times$ 10$^6$ K and V$_{\rm wind}$
$\approx$ 200 - 500 km s$^{-1}$ is driven off if L/L$_E$ $\ga$ 0.08
(e.g., Balsara \& Krolik 1993). The wind is directed along the minor
axis of the {\em accretion} disk and is not necessarily perpendicular to
the {\em galactic} disk.
\item[3.] {\bf Jet-driven thermal winds.} AGN-driven jets entrain and
heat gas on kiloparsec scales. The internal energy densities of these
loosely collimated jets may be dominated by thermal X-ray-emitting
plasma. The extended soft X-ray emission from these objects should be
roughly cospatial with the large-scale radio emission, as seen in some
active galaxies (e.g., Colbert et al. 1996, 1998). The
transport of energy and momentum by these ``mass-loaded'' jets may be
able to power the narrow-line regions of Seyferts (e.g., Bicknell et al. 1998).
\end{itemize}

In the rest of this paper, we aim to determine which one(s) of these
mechanisms is(are) responsible for the wide-angle outflows seen in
AGNs. In \S 2, we describe a survey our group has been conducting over
the years on nearby AGNs. In \S 3, we summarize the general trends
that we see in our data and address the issues of the ionization
source of the line-emitting material and the energy source of the
outflows. In \S 4, we discuss the recent results on three objects
representative of our sample: Circinus, NGC~2992, and NGC~4388. We
summarize our conclusions and discuss future avenues of research in \S
5.

\section{Description of Survey}

Over the past ten years, our group has been conducting an optical
survey of nearby active and starburst galaxies combining Fabry-Perot
imaging and long-slit spectrophotometry with radio and X-ray data to
track the energy flow of galactic winds through the various gas
phases.  The complete spatial and kinematic sampling of the
Fabry-Perot data is ideally suited to study the complex and extended
morphology of the warm line-emitting material which is associated with
the wind flow. The radio and X-ray data complement the Fabry-Perot
data by probing the relativistic and hot gas components, respectively.

Our sample contains twenty active galaxies with known galactic-scale
outflows. Half of them are Seyfert galaxies and the rest are starburst
galaxies (not discussed here; the outflows in these objects are
starburst-driven winds). All of these objects are nearby ($z$ $<$ 0.01
to provide a spatial scale of $<$ 200 pc arcsec$^{-1}$) and have
line-emitting regions that extends more than 30$\arcsec$. So far, the
results have been published for about a dozen of these objects. Table
1 lists the main papers which have come out from our survey.

\begin{table}[!b]
  \begin{center}
    \caption{Some key publications}  
    \label{tab:1}
    \begin{tabular}{ll}\hline\hline
      Galaxy & Reference \\ \hline
      M 51    & Cecil (1988) \\
      M 82    & Bland \& Tully (1988) \\
              & Shopbell \& Bland-Hawthorn (1998) \\
      NGC 1068& Cecil et al. (1990) \\
              & Bland-Hawthorn et al. (1991)\\
              & Sokolowski et al. (1991)\\
              & Cecil et al. (2001b) \\
      NGC 2992& Veilleux et al. (2001) \\
      NGC 3079& Veilleux et al. (1994)\\
              & Veilleux et al. (1995)\\
              & Veilleux et al. (1999a)\\
              & Cecil et al. (2001a)\\
      NGC 3516& Veilleux et al. (1993)\\
      NGC 4258& Cecil et al. (1992)\\
              & Cecil et al. (1995)\\
              & Cecil et al. (1995)\\
              & Cecil et al. (2000)\\
      NGC 4388& Veilleux et al. (1999a)\\
              & Veilleux et al. (1999b)\\
      NGC 6240& Bland-Hawthorn et al. (1991)\\
      Circinus& Veilleux \& Bland-Hawthorn (1997)\\
      \hline\hline
    \end{tabular}
  \end{center}
\end{table}

\section{General Results}

Evidence for loosely collimated winds are detected in several
AGNs. Approximately 75\% of the AGNs in our (admittedly biased) sample
harbor wide-angle outflows rather than collimated jets. These outflows
typically show the following optical properties:

\begin{itemize}

\item[$\bullet$] The optical winds are often lopsided and sometimes tilted with
respect to the polar axis of the host galaxy (e.g., NGC~4388).
\item[$\bullet$] The solid angle subtended by these winds, $\Omega_W/4\pi$ 
$\approx$ 0.1 -- 0.5.
\item[$\bullet$] The radial extent of the line-emitting material involved in the outflow, 
$R_W$ = 1 -- 5 kpc.
\item[$\bullet$] The outflow velocity of the line-emitting material, $V_W$ = 100 -- 1500 km s$^{-1}$ regardless of the escape velocity of the
host galaxy. Until recently the record holder was NGC~3079, where outflow
velocities in excess of 1500 km s$^{-1}$ are directly measured (Veilleux et 
al. 1994). But the outflow velocities in NGC~1068 are now known to be far 
larger than this value (see Cecil, these proceedings, and Cecil et al. 2001b). 
\item[$\bullet$] The dynamical time scale, $t_{\rm dyn}$ $\approx$ $R_W/V_W$ = $10^6 - 10^7$ years.
\item[$\bullet$] The ionized mass involved in the outflow, 
$M$ = 10$^5$ -- 10$^7$ M$_\odot$, a relatively small fraction of the total 
ISM in the host galaxy. 
\item[$\bullet$] The ionized mass outflow rate, d$M$/d$t$ $\approx$ $M$/$t_{\rm
dyn}$ = 0.1 -- 1 $M_\odot$ yr$^{-1}$ $>$ d$M_{\rm acc}$/d$t$,
the mass accretion rate necessary to fuel the AGN.
\item[$\bullet$] The kinetic energy of the outflowing ionized material, $E_{\rm
kin}$ = 10$^{53}$ -- 10$^{55}$ ergs, taking into account both the bulk and 
``turbulent'' (spectrally unresolved) motions. This mechanical energy is 
equivalent to that of $\sim$ 10$^2$ -- 10$^4$ Type II SNe. 
\item[$\bullet$] The kinetic energy rate of the outflowing ionized material,
d$E_{\rm kin}$/d$t$ $\approx$ $E_{\rm kin}/t_{\rm dyn}$ = few
$\times 10^{39} - 10^{42}$ ergs s$^{-1}$.
\item[$\bullet$] Evidence for entrainment of (rotating) disk material is seen
in some objects (e.g., Circinus, NGC~2992, NGC~3079).
\item[$\bullet$] The source of ionization of the line-emitting material taking
part in these outflows is diverse.  Pure photoionization by the AGN can 
explain the emission-line ratios in NGC~3516 and NGC~4388. Shock ionization 
is probably contributing in Circinus and NGC~2992.
\item[$\bullet$] Starburst-driven winds are not common among AGNs (a possible
exception is NGC~3079; Veilleux et al. 1994; Cecil et
al. 2001a). Galactic winds are roughly aligned with galaxy-scale (several kpc) 
radio emission, sometimes encompassing what appears to be poorly collimated
radio ``jets''. These mass-loaded ``jets'' are the probable driving
engine for these winds. In most cases, however, torus-driven winds cannot 
formally be ruled out.
\end{itemize}

\section{Case Studies of Three Nearby AGNs}

In this section, we discuss the recent results on Circinus, NGC~2992,
and NGC~4388, three fairly representative objects of our sample. While
the morphology and kinematics of the optical outflows in these objects
are very different, they appear fundamentally to be driven by the same
process, namely entrainment along poorly collimated ``jets''. For more
information on these objects, the reader should refer to the original
papers: Veilleux \& Bland-Hawthorn (1997), Veilleux et al. (2001), and
Veilleux et al. (1999a,b), respectively.

\subsection{Circinus}

The Circinus galaxy is the nearest ($\sim$ 4 Mpc) Seyfert 2 galaxy
known, and is therefore ideally suited for detailed studies of
AGN-driven outflows. Recent maps (Elmouttie et al. 1995, 1998) have
resolved spectacular radio lobes centered on the nucleus, and
extending more than 90\arcsec\ ($\sim$ 2 kpc) on either side of the
galaxy disk (PA $\approx$ --60$^\circ$).

\subsection{Morphology and Kinematics of the Line-Emitting Material}

Our Fabry-Perot data on Circinus (Fig. 1) reveal a complex of ionized
filaments extending radially from the nucleus out to distances of 1
kpc.  The most striking [O~III] feature extends along position angle
$\sim$ --50$^\circ$ spanning a distance of $\sim$ 25 -- 45\arcsec\
(500 -- 900 pc) from the nucleus. The lateral extent of this filament
is near the limit of our resolution ($\sim$ 1\farcs 5 after
smoothing). This narrow feature is also visible at H$\alpha$ but with
a lower contrast. The gas at this location is highly ionized with a
[O~III] $\lambda$5007/H$\alpha$ flux ratio typically larger than
unity. Extrapolation of this filament to smaller radii comes to within
1\arcsec\ (20 pc) of the infrared nucleus (Marconi et al. 1995),
suggesting a nuclear (most likely AGN) origin to this feature.

The most spectacular feature in the H$\alpha$ data is the hook-shaped
filament which extends to 40\arcsec\ (800 pc) west of the nucleus
(Fig. 2).  Such features are commonly observed in HH objects although
they have never been seen on galactic scales. Additional morphological
evidence for outflow exists in the northern portion of our data.  The
[O~III] emission along PA $\approx$ --20$^\circ$ forms a broad
filamentary `finger' or jet that points back to the nucleus. A knot is
present at the tip of this `finger', 25\arcsec\ from the nucleus.
Bright H$\alpha$ emission is also visible near this position, the
southern portion of which forms a wide ($\sim$ 8\arcsec\ ) arc
resembling a bow shock. The arc is pointing in the downstream
direction consistent with being produced by a collimated jet.
\begin{figure}
  \includegraphics[width=0.75\columnwidth]{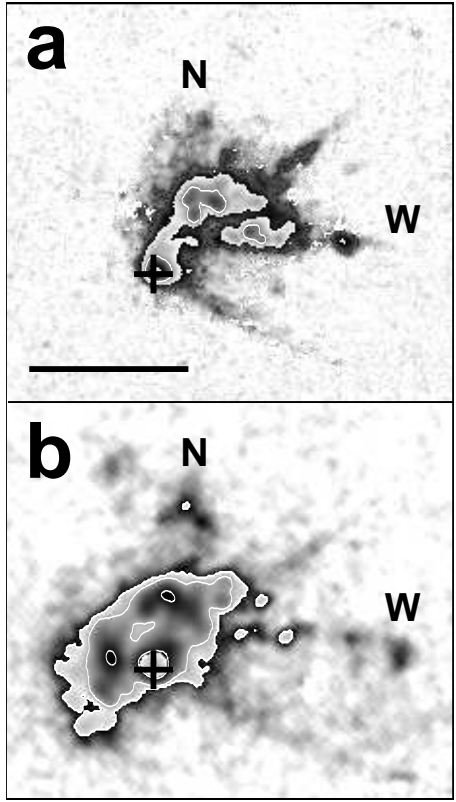}
  \caption{ Line flux images of the Circinus galaxy: $a$, [O~III]
$\lambda$5007 and $b$, blueshifted (between --150 and 0 km s$^{-1}$) H$\alpha$.
North is at the top and west to the right.  The position of the
infrared nucleus (Marconi et al. 1995) is indicated in each image by a
cross. The spatial scale, indicated by a horizontal bar at the bottom
of the [O~III] image, is the same for each image and corresponds to
$\sim$ 25\arcsec\, or 500 pc for the adopted distance of the Circinus
galaxy of 4 Mpc.  The minor axis of the galaxy runs along PA $\approx$
--60$^\circ$. }
\end{figure}
\begin{figure}
  \includegraphics[width=0.75\columnwidth]{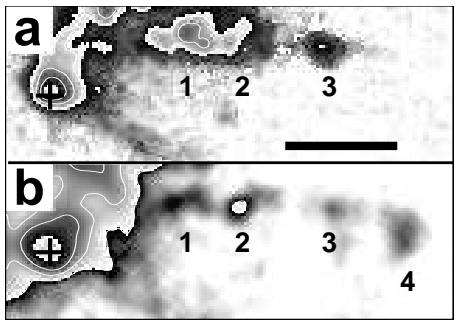}
  \caption{ Line flux images of the western hook-shaped filament: $a$,
[O~III] $\lambda$5007 and $b$, blueshifted (between --150 and 0 km
s$^{-1}$) H$\alpha$. The position of the infrared nucleus (Marconi et
al. 1995) is indicated in each image by a cross.  The orientation is
the same as in Fig. 1, but the horizontal bar at the bottom of the
[O~III] image now corresponds to $\sim$ 250 pc.  
 }
\end{figure}

The kinematics derived from the [O~III] Fabry-Perot data and
complementary long-slit spectra bring credence to the nuclear outflow
scenario.  Non-gravitational motions are observed throughout the
[O~III] cone, superposed on a large-scale velocity gradient caused by
galactic rotation along the major axis of the galaxy (PA$_{\rm maj}$
$\approx$ 30$^\circ$; Freeman et al. 1977).  Sudden velocity gradients
are seen near the positions of the bright [O~III] and H$\alpha$ knots.

\subsubsection{Source of Ionization}

The motions observed across the ionization cone are highly supersonic,
so high-velocity ($V_s$~$\ga$~100~km~s$^{-1}$) shocks are likely to
contribute to the ionization of the line emitting gas.  Large
variations of the line ratios are sometimes observed within a single
knot (Fig. 3). The enhanced [N~II]/H$\alpha$, [S~II]/H$\alpha$, and
[O~III]/H$\beta$ ratios in knots 1 and 2 fall near the range produced
by the high velocity photoionizing radiative shocks of Dopita \&
Sutherland (1995).  Photoionization purely by the active nucleus of a
mixture of ionization and matter-bounded clouds whose relative
proportions vary with position in the galaxy may also provide another
explanation for the abrupt changes of excitation in the filaments
(e.g., Binette, Wilson, \& Storchi-Bergmann 1996).

\begin{figure}
  \includegraphics[width=0.8\columnwidth]{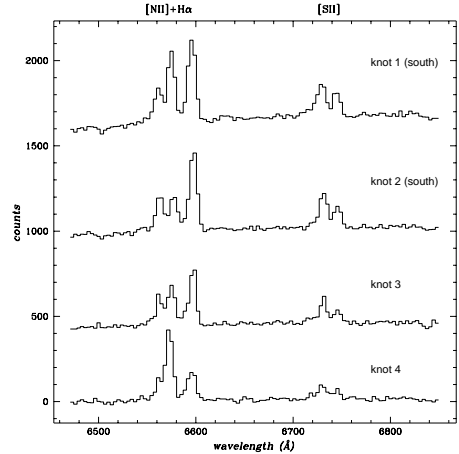}
  \caption{Binned spectra at four positions along the slit through the nucleus
corresponding to knots 1$-$4 in Fig. 2. Each spectrum has been offset by 400
counts. Knots 1 and 2 are binned along the slit over 6\arcsec; knots 3
and 4 are binned over 4\arcsec. Note the greatly enhanced [N~II]/H$\alpha$
ratio along the jet, except on the bow shock at knot 4. 
}
\end{figure}

\subsubsection{Nature of the Outflow}

The mass of ionized gas involved in this outflow is fairly modest,
$\sim$ few $\times$ 10$^4$ X$^{-1}$ n$_{e,2}^{-1}$ M$_\odot$ where X
is the fraction ($<$ 1) of oxygen which is doubly ionized and
N$_{e,2}$ is the electron density in units of 10$^2$ cm$^{-3}$. The
total kinetic energy ($\sim$ 10$^{53}$ n$_{e,2}^{-1}$ ergs) lies near
the low energy end of the distribution for wide-angle events observed
in nearby galaxies. The morphology and velocity field of the filaments
suggest that they represent material expelled from the nucleus
(possibly in the form of ``bullets'') or entrained in a wide-angle
wind roughly aligned with the radio ``jet'' and the polar axis of the
galaxy. The complex morphology of the outflow in the Circinus galaxy
is unique among active galaxies. The event in the Circinus galaxy may
represent a relatively common evolutionary phase in the lives of
gas-rich active galaxies during which the dusty cocoon surrounding the
nucleus is expelled by the action of jet or wind phenomena.

\subsection{NGC~2992}

The presence of a galactic-scale outflow in NGC~2992 has been
suspected for several years, based on the morphology of the radio
emission (Ward et al. 1980; Hummel et al. 1983) and more recently the
X-ray emission (Colbert et al. 1998). In the optical, the line
emission from the outflow is severely blended with line emission from
the galactic disk. The complete two-dimensional coverage of our
Fabry-Perot data is therefore critical to disentangle the
material associated with the wind from the gas in rotation in the
galactic disk.

\subsubsection{Morphology and Kinematics of the Outflowing Line-Emitting 
Material}

The distribution of the H$\alpha$-emitting gas in the disk and outflow
components is shown in Figure 4. The outflow component (Fig. 4$d$) is
distributed into two wide cones which extend up to $\sim$ 18$\arcsec$
(2.8 kpc) from the peak in the optical continuum map.  Both cones have
similar opening angles of order 125$\arcdeg$ -- 135$\arcdeg$.  The
bisectors of each of the cones coincide with each other and lie along
P.A. $\approx$ 116$\arcdeg$, or almost exactly perpendicular to the
kinematic major axis of the inner disk (P.A. $\approx$
32$\arcdeg$). This strongly suggests that the axis of the bicone is
perpendicular to the galactic disk, and that the material in the SE
cone is emerging from under the galaxy disk while the material in the
NW cone is emerging from above the disk.

The outflow on the SE side of the nucleus is made of two distinct
kinematic components interpreted as the front and back walls of a
cone. The azimuthal velocity gradient in the back-wall component
reflects residual rotational motion which indicates either that the
outflowing material was lifted from the disk or that the underlying
galactic disk is contributing slightly to this component.  A single
outflow component is detected in the NW cone.

\begin{figure}
  \includegraphics[width=\columnwidth]{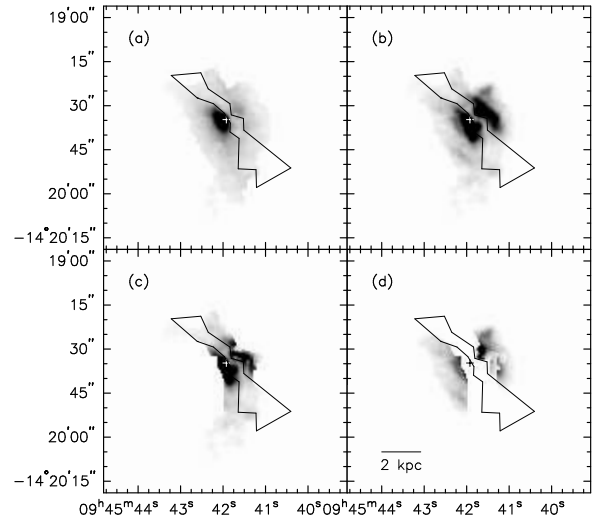}
  \caption{Distribution of the H$\alpha$ and continuum emission in
NGC~2992.  ($a$) Continuum emission at rest wavelength $H\alpha$,
($b$) total H$\alpha$ emission, ($c$) H$\alpha$ emission from the disk
component, ($d$) H$\alpha$ emission from the outflow component. The
position of the radio nucleus (``+'') and the extent of the dust lane
are indicated on each of these panels. North is at the top and east to
the left. Note the absence of a disk component in the east
quadrant.}
\end{figure}

\begin{figure}
  \includegraphics[width=\columnwidth]{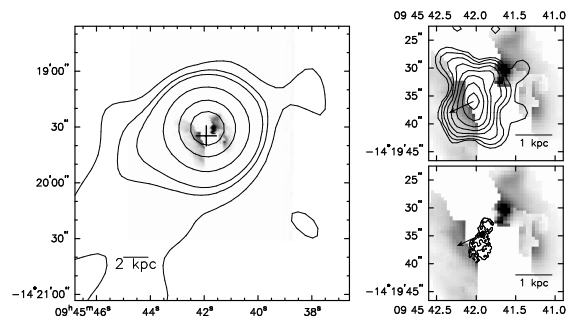}
  \caption{Contour maps of the X-ray emission (left panel: ROSAT/HRI
data from Colbert et al. 1998) and 6-cm continuum emission (upper
right panel: VLA C-configuration data from Colbert et al. 1996 with
3$\arcsec$ uniform weighting; lower right panel: VLA A-configuration
data from Ulvestad \& Wilson 1984) superimposed on the total H$\alpha$
emission from NGC~2992 (grey-scale). A cross (``+'') in the main
figure indicates the position of the radio nucleus. The arrows in the
right panels mark the direction of the one-sided 90$\arcsec$ (13.5
kpc) radio extension along P.A. $\sim$ 100$\arcdeg$ -- 130$\arcdeg$,
i.e. close to the axis of the biconical outflow. Same orientation as
Fig. 4.}
\end{figure}

\subsubsection{Source of the Outflow}

A biconical outflow model with velocities ranging from 50 to 200
km~s$^{-1}$ and oriented nearly perpendicular to the galactic disk can
explain most of the data.  The broad line profiles and asymmetries in
the velocity fields suggest that some of the entrained line-emitting
material may lie inside the biconical structure rather than only on
the surface of the bicone. The mass involved in this outflow is of
order $\sim$ 1 $\times$ 10$^7$ n$_{e,2}^{-1}$ M$_\odot$, and the bulk
and ``turbulent'' kinematic energies are $\sim$ 6 $\times$ 10$^{53}$
n$_{e,2}^{-1}$ ergs and $\sim$ 3 $\times$ 10$^{54}$ n$_{e,2}^{-1}$
ergs, respectively.  The faint X-ray extension detected by Colbert et
al. (1998) and shown in Figure 5 lies close to the axis of the outflow
bicone, suggesting a possible link between the hot and warm gas phases
in the outflow. The position angles of the optical outflow bisector
and the long axis of the ``figure-8'' radio structure shown in Fig. 5
differ by $\sim$ 25$\arcdeg$ -- 35$\arcdeg$, but the one-sided
13.5-kpc radio extension detected by Ward et al. (1980) and Hummel et
al. (1983) lies roughly along the same direction as the mid-axis of
the SE optical cone (P.A. $\sim$ 100$\arcdeg$ -- 130$\arcdeg$; arrows
mark this direction in the right panels of Fig. 5).  The most likely
energy source of the optical outflow is a hot bipolar thermal wind
powered on sub-kpc scale by the AGN and diverted along the galaxy
minor axis by the pressure gradient of the ISM in the host galaxy.
The data are not consistent with a starburst-driven wind or a
collimated outflow powered by radio jets.

\subsection{NGC~4388}

NGC~4388 was the first Seyfert galaxy discovered in the Virgo cluster
(Phillips \& Malin 1982). It is plunging edge-on at 1500 km s$^{-1}$
through the center of the Virgo cluster and experiencing the effects
of ram-pressure stripping by the densest portions of the intracluster
medium (ICM; e.g., Cayatte et al. 1994). Nuclear activity has been
detected at nearly all wavelengths. The morphology of the nuclear
radio source suggests a collimated AGN-driven outflow (Stone et
al. 1988; Falcke, Wilson, \& Simpson 1998).  At optical wavelengths,
NGC 4388 has been known for some time to present extended line
emission (e.g., Pogge 1988; Corbin, Baldwin, \& Wilson 1988 and
references therein). A rich complex of ionized gas extends both along
the disk of the galaxy and up to 50$\arcsec$ (4 kpc) above that plane.
The extraplanar gas component has considerably higher ionization than
the disk gas, and appears roughly distributed into two opposed
radiation cones that emanate from the nucleus (Pogge 1988; Falcke et
al. 1998). Detailed spectroscopic studies strongly suggest that this
component is mainly ionized by photons from the nuclear continuum
(Pogge 1988; Colina 1992; Petitjean \& Duret 1993).

\subsubsection{Morphology and Kinematics of the Extraplanar Line-Emitting 
Material}

Figure 6 shows the distribution of the
line-emitting material in this galaxy derived from our Fabry-Perot
data. We confirm the existence of a rich complex of highly ionized gas
that extends $\sim$ 4 kpc above the disk of this galaxy.
Low-ionization gas associated with star formation is also present in
the disk. Evidence for bar streaming is detected in the disk component
and is discussed in Veilleux et al. (1999a,b).  Non-rotational
blueshifted velocities of 50 $-$ 250 km s$^{-1}$ are measured in the
extraplanar gas north-east of the nucleus.  The brighter features in
this complex tend to have more blueshifted velocities. A redshifted
cloud is also detected 2 kpc south-west of the nucleus.  The total
mass and kinetic energy of the extraplanar gas are $\sim$ 4 $\times$
10$^5$ M$_\odot$ and $E_{\rm kin}$ = $E_{\rm bulk}$ + $E_{\rm turb}$
$\ga$ 1 $\times$ 10$^{53}$ ergs, respectively.

\begin{figure}
  \includegraphics[width=0.9\columnwidth]{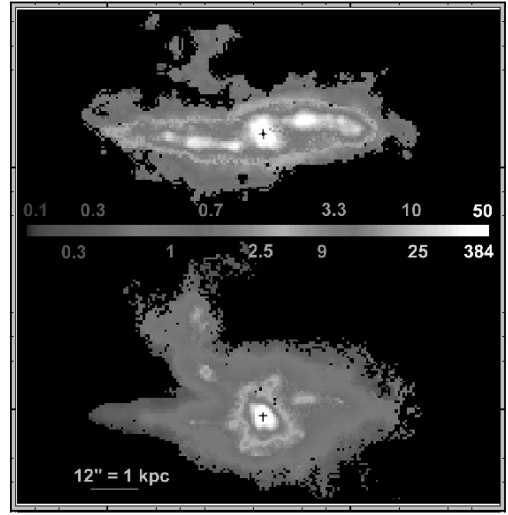}
  \caption{ The distributions of the line emission in NGC 4388. (top)
H$\alpha$; (bottom) [O~III] $\lambda$5007. North
is at the top and East to the left. The spatial scale, indicated by a
horizontal bar at the bottom of the image, corresponds to 12$\arcsec$,
or 1 kpc for the adopted distance of 16.7 Mpc. The optical continuum
nucleus is indicated in each panel by a cross.
}
\end{figure}

\subsubsection{Geometry of the Outflow}

The velocity field of the extraplanar gas of NGC~4388 appears to be
unaffected by the inferred supersonic (Mach number $M$ $\approx$ 3)
motion of this galaxy through the ICM of the Virgo cluster.  This is
because the galaxy and the high-$\vert$z$\vert$ gas lie behind a Mach
cone with opening angle $\sim$ 80$\arcdeg$ (see Fig. 7).  The shocked
ICM that flows near the galaxy has a velocity of $\sim$ 500 km
s$^{-1}$ and exerts insufficient ram pressure on the extraplanar gas
to perturb its kinematics.  In Veilleux et al. (1999b), we consider
several explanations for the velocity field of the extraplanar
gas. Velocities, especially blueshifted velocities on the N side of
the galaxy, are best explained as a bipolar outflow which is tilted by
$>$12$\arcdeg$ from the normal to the disk.  The observed offset
between the extraplanar gas and the radio structure may be due to
buoyancy or refractive bending by density gradients in the halo gas.

\begin{figure}
  \includegraphics[width=0.9\columnwidth]{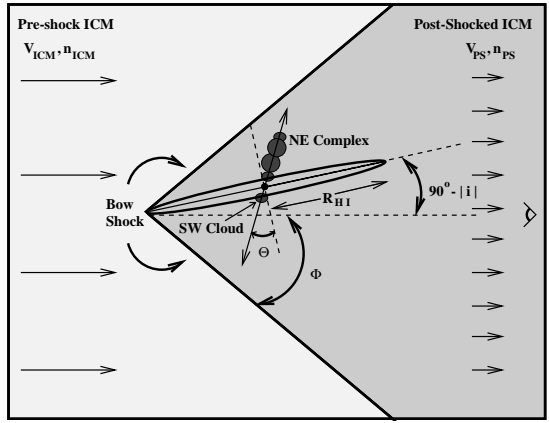} 
  \caption{ Geometry of
  the ICM -- NGC 4388 interaction. The values of the parameters
  indicated on this figure and discussed in Veilleux et al. (1999b)
  are: $i$ = -- 78$\arcdeg$, $\phi$ = 40$\arcdeg$, $\theta$ $>$
  12$\arcdeg$, $R_{\rm H~I}$ = 10 kpc, $V_{\rm ICM}$ = 1,500 km
 s$^{-1}$ , $n_{\rm ICM}$ $\sim$ 10$^{-4}$ cm$^{-3}$, $V_{\rm ps}$ =
  500 km s$^{-1}$, $n_{\rm ps}$ $\sim$ 3 $\times$ 10$^{-3}$
  cm$^{-3}$. The observer is located on the right in the same plane as
 the figure and at a distance of 16.7 Mpc.}
\end{figure}

\section{Summary \& Future Work}

Over the past ten years, our group has obtained high-quality optical
Fabry-Perot and long-slit spectrophotometry of a large sample of
active galaxies. Many of these sources harbor wide-angle AGN-driven
winds which can be of importance in the chemical and thermal evolution
of the host galaxies.  The high level of sophistication of recent
hydrodynamical simulations (e.g., Suchkov et al. 1994; Strickland \&
Stevens 2000) has provided the theoretical basis to interpret our data
and to predict the evolution and eventual resting place (disk, halo,
or intergalactic medium) of the outflowing material. In the coming
years, new ground-based observational techniques (e.g., tunable
narrow-band filters, nod and shuffle techniques; Bland-Hawthorn 2000)
and spacecrafts (e.g., Chandra and XMM) will allow us to put even
stronger constraints on the influence of AGN-driven winds in nearby
galaxies.  These data will provide a critical local baseline for
future surveys of high-redshift sources.

\acknowledgements S.V. is grateful for partial support of this research
by a Cottrell Scholarship,
NASA/LTSA grant NAG 56547, and NSF/CAREER grant AST-9874973.


\end{document}